\documentclass[twocolumn,showpacs,preprintnumbers,amsmath,amssymb]{revtex4}
\usepackage{graphicx}% Include figure files
\usepackage{dcolumn}% Align table columns on decimal point
\usepackage{bm}% bold math

\newcommand {\be}{\begin{equation}}
\newcommand {\ee}{\end{equation}}
\newcommand {\bea}{\begin{eqnarray}}
\newcommand {\eea}{\end{eqnarray}}
\newcommand {\bem}{\begin{displaymath}}
\newcommand {\eem}{\end{displaymath}}
\newcommand {\f}{\frac }
\newcommand\D[2]{\frac{\partial #1}{\partial #2}}

\newcommand {\p}{\partial}

\begin{document}

\preprint{ }

\title{The effects of superconductor-stabilizer interfacial resistance on quench of a pancake coil made out of coated conductor}% Force line breaks with \\
\author{ G. A. Levin, W. A. Jones } 
\affiliation{Air Force Research Laboratory, Propulsion Directorate, Wright-Patterson Air Force Base, OH 45433 USA}
\author{ K. A. Novak}
\affiliation{Department of Mathematics, Air Force Institute of Technology, Wright-Patterson Air Force Base, OH 45433 USA}
\author{P. N. Barnes}
\affiliation{Air Force Research Laboratory, Propulsion Directorate, Wright-Patterson Air Force Base, OH 45433 USA}

\date{\today}% It is always \today, today,
             %  but any date may be explicitly specified

\begin{abstract}
We present the results of numerical analysis of normal zone propagation in a stack of $YBa_2Cu_3O_{7-x}$  coated conductors which imitates a pancake coil. Our main purpose is to determine whether the quench protection quality of such coils can be substantially improved by increased contact resistance between the superconducting film and the stabilizer.  We show that with increased contact resistance the speed of normal zone propagation increases, the detection of a normal zone inside the coil becomes possible earlier, when the peak temperature inside the normal zone is lower, and stability margins shrink. Thus, increasing contact resistance may become a viable option for improving the prospects of coated conductors for high $T_c$ magnets applications. 
\end{abstract}
%\keywords{Suggested keywords}%Use showkeys class option if keyword
                              %display desired
\maketitle

\section{\label{sec:introduction}Introduction\protect}

High temperature superconducting wires such as $YBa_2Cu_3O_{7-x}$  ($YBCO$)  coated conductors~\cite{Rup} have a potential for a number of important applications. All of them, and especially the high field magnets, require effective quench detection and protection systems\cite{Bray}. Currently manufactured coated conductors are characterized by very low speed of normal zone propagation (NZP)~\cite{Sch,Song,Sumption}.  This makes it difficult to promptly detect the onset of a quench with the result that in a poorly cooled (adiabatic) coil the temperature of the hot spot will rise above the safe limit before the quench protection system is engaged, irreversibly damaging the conductor. Unlike the conventional low$-T_c$ wires in which a quench can be easily triggered by a small amount of energy, the coated conductors are substantially more stable.   Thus, the quench in devices made out of coated conductors has a potential to become a Black Swan event -- a rare, unpredictable occurrence with catastrophic consequences. 
\begin{figure}
\includegraphics{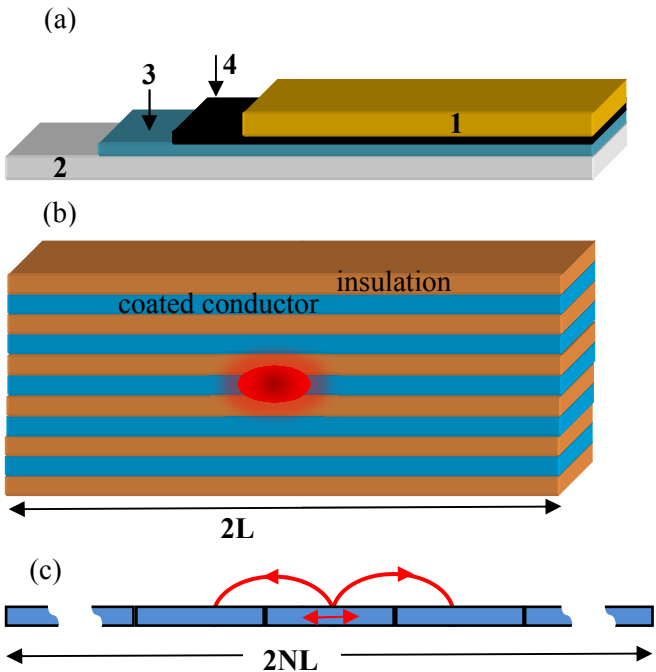}
\\
\caption{\label{fig:} (a) A sketch of a coated conductor (not to scale). A thin superconducting ($YBCO$ ) film (3) is deposited on metal substrate (2).  The resistive interface (4) segregates copper stabilizer (1) and $YBCO$ film. (b) A stack of $N$ coated conductors (each similar to that in (a)) segregated by electric insulation with finite thermal conductivity. The stack as a whole is thermally insulated from the environment (adiabatic). A blob in the center of the stack indicates a nucleated normal zone. (c) A model described by Eq. (2) corresponds to all conductors in the stack being connected in series. The arrows indicate the lateral heat transfer (along the conductor), as well as the thermal coupling over the finite distance $2L$ which corresponds to the heat exchange in the transverse direction, between the neighboring conductors in the stack shown in (b).}
\end{figure}

A promising approach to improving the quench protection quality of coated conductors is to increase the contact resistance between the superconducting film and stabilizer. In Refs. \cite{ASC,L1} it was shown, using a model of a linear conductor, that the NZP speed can be significantly increased by placing a high resistance interface between the YBCO film and copper stabilizer. The effect of increased contact resistance on the superconducting wire has been studied in the past. A practical reason for experimental and theoretical work in this area was a potential for reduction of coupling AC losses in the low $T_c$ wires. A strong decline in stability margins which results from large contact resistance and technological difficulties of making such resistance consistently uniform prevented the incorporation of this approach into practice. Earlier patents on the subject are difficult to locate. An extensive, although perhaps not entirely comprehensive, list of references can be found in the reviews \cite{G1,G2}. 

The property of high contact resistance to increase the NZP speed was not addressed in this earlier research probably because increasing the NZP speed in low $T_c$ wires was not necessary as it was sufficiently high to assure adequate quench protection. The advent of coated conductors has changed the priorities. Coated conductors are much more stable than the low $T_c$ wires, so that the reduction of stability margins accompanied by increasing NZP speed and with it an earlier detection of quench may be desirable. The sequential stages of the manufacturing process of coated conductors are also conducive to incorporating an additional step of adding a thin, high resistance interface between the YBCO film and stabilizer.

Here we present an analysis of a model of quench in a pancake coil made out of coated conductor. It should be noted that increasing NZP speed is not in and of itself our goal. The propagating quench builds up the electric potential difference across the coil. Typically a quench protection system is engaged when this potential exceeds a certain threshold. Therefore, a goal of a quench detection system is to reliably detect quench in the shortest amount of time, thus minimizing the peak temperature inside the normal zone.  The effect of increased contact resistance on quench is not straightforward. The NZP speed increases, but the power dissipation at the front of the propagating NZ also increases. Thus, in the worst case scenario of adiabatic quench these two effects compete with each other. We show that the increased contact resistance has an overall positive effect by increasing the rate at which the potential drop across the coil increases and, at the same time, reducing the rate at which the peak temperature inside the normal zone rises with the potential drop across the coil. Thus, at the moment of quench detection the peak temperature inside the coil will be lower, which leaves a greater safety margins for the quench protection system.

\section{\label{sec:model}Model of quench propagation in a pancake coil
\protect}

In~\cite{PRE} it was shown that the 3D equation of heat conduction in a thin tape-like composite wire, Fig. 1(a), can be reduced to a 2D (planar) equation if the heat flux from the surface does not greatly exceed $1\;W/cm^2$. Then, the temperature variation across the thickness of the tape constitutes a fraction of a degree, much smaller than the variation of temperature along the wire. The 1D approximation is valid as long as the thermal diffusion length is greater or comparable to the conductor width. In coated conductors this condition is met.
 
A quench in pancake coil can be adequately modeled by a stack of coated conductors segregated by layers of insulation as shown in Fig. 1(b). One can view such a stack as part of the straight section of a race-track pancake coil. Incidentally, the stacks of coated conductors were used in experimental studies of quench reported in \cite{Sumption}. The model of the coated conductor itself is the same as in Ref. \cite{L1}. We take into account four constituencies of the coated conductor: stabilizer, substrate, superconducting YBCO film, and infinitesimally thin interface between YBCO and stabilizer which accounts for the finite contact resistance, Fig. 1(a). 

The system of equations describing the heat exchange among the coated conductors in the stack takes form:
\bea \label{E:heat_equation_1}
C\f{\p T_i}{\p t}-\f{\p}{\p x}\left (K\f{\p T_i}{\p x}\right ) =
Q_i - K_{\perp}(T_i(x)-T_{i+1}(x)) \\\nonumber
-K_{\perp}(T_i(x)-T_{i-1}(x)).
\eea
The conductors in the stack, each of length $2L$, are enumerated from $i=1$ to $N$. The dimensionality of Eq.~\eqref{E:heat_equation_1} is $[W/cm^2]$. Here $K_{\perp}=k_{ins}/2d_{ins}\;[Wcm^{-2}K^{-1}]$ is the heat conductance across the insulation layer segregating the neighboring turns of the coil. We consider that both sides of the conductors are wrapped in an insulating film of thickness $d_{ins}$. Equation (1) describes the worst case scenario, from the quench protection point of view,  of fully adiabatic case with no heat exchange between the coil and the environment.  $C=c_1d_1+c_2d_2+2c_{ins}d_{ins}\;[Jcm^{-2}K^{-1}]$ is the combined heat capacity of the coated conductor plus insulation per unit area, $K=k_1d_1+k_2d_2\;[W\;K^{-1}]$ is the effective in-plane thermal conductance (the in-plane thermal conductance of the insulation can be neglected). The subscripts 1, 2, and $ins$ refer to the stabilizer, substrate, and insulation, respectively. The thicknesses of the stabilizer and substrate are denoted as $d_1$ and $d_2$ respectively. Here $c_1,c_2,c_{ins}, k_1,k_2$, and $k_{ins}$ are the respective material constants. $Q_i=\int_{-d_2}^{d_1}q_i(z)dz$ is the density of the internal heat sources integrated over the thickness of the conductor. %$K_0$ is the heat transfer coefficient across the insulation on the %surface of the wire and $T_0$ is the coolant (ambient) temperature. 

Since all conductors in the stack shown in Fig. 1(b) are parts of one conductor and carry the same current, it is convenient to reformulate the problem of N coupled equations as one equation similar to that analyzed in Ref.\cite{L1}. We will consider the conductors in the stack to be connected in series, as shown in Fig. 1(c). The total length of the resulting conductor is $2NL$. 
The difference between the quench propagation in a single linear conductor and that in a coil is the thermal coupling between the turns which in our case translates into thermal coupling between the points separated by a finite distance $2L$, as shown in Fig. 1(c). Thus, the system of Eqs.~\eqref{E:heat_equation_1} can be
rewritten as 
\bea \label{E:heat_equation_2}
C\f{\p T}{\p t}-\f{\p}{\p x}\left (K\f{\p T}{\p x}\right ) =
Q - K_{\perp}(T(x)-T(x+2L)) \\\nonumber
-K_{\perp}(T(x)-T(x-2L)).
\eea
Any pancake coil, round or race track shape, can be modeled by a similar equation, except that the finite distances between the coupled points will vary along the wire. 

The second equation defined for a conductor shown in Fig. 1(c) describes the instantaneous redistribution of current between the superconducting film and stabilizer and is determined by the condition of charge conservation 
\begin{equation}\label{E:charge_conservation_1}
\D{J_1}{x}=-\f{V_1  -V_s  }{\bar {R}}, 
\end{equation}
where $J_1$ is the linear density of current ($A\;cm^{-1}$) flowing through the stabilizer. $V_1$ and $V_s$ are the local electric potentials of the stabilizer and superconductor, respectively and $\bar {R}\;[\Omega\;cm^2]$ is the resistance of the unit area of the interface (contact resistance). This condition can also be used in the form
\begin{equation}\label{E:charge_conservation_2}
\D{}{x}\left (\bar{R}\D{J_1}{x}\right )= E_1 - E_s.
\end{equation}
Here $E_1  =-\p V_1  /\p x$ and $E_s =-\p V_s/\p x$ are the electric fields in the stabilizer and superconductor, respectively, and we do not assume that $\bar {R}$ is uniform. 

The integrated area density of heat sources in~\eqref{E:heat_equation_2} takes the form
\begin{equation}\label{E:integrated_internal_heat_sources}
Q=\f{d_1}{\rho_1}E_1^2  +\f{(V_1  -V_s)^2}{\bar {R}} 
+J_sE_s  ,
\end{equation}
where $J_s =J-J_1$ is the density of current in the superconductor and $J=const$ is the total transport current density. 

The constituent relationship between current density and electric field for a superconductor will be used in the same form as in ~\cite{L1} 
\begin{equation}\label{E:piecewise_field}
E_s=\begin{cases}
\f{\rho_s}{d_s} (J_s-J_c),&\text{if } J_s>J_c \\
0,&\text{if } J_s\le J_c
\end{cases},
\end{equation}
where $\rho_s$ is the normal state resistivity and $d_s$ the thickness of the superconducting film. Hereafter, we will adopt a linear temperature dependence  of $J_c(T)$~\cite{Ekin}:
\begin{equation}\label{E:linear_temperature_dependence}
J_c(T)=a(T_c-T);\;\; T<T_c.
\end{equation}
For the stabilizer, the conventional Ohmic relationship will suffice at all temperatures:
\begin{equation}
E_1=\f{\rho_1}{d_1}J_1.
\end{equation}
In the normal state the resistance of YBCO film is much greater than that of the stabilizer,
\begin{equation}
\f{\rho_s}{d_s}\gg \f{\rho_1}{d_1}.
\end{equation}

The final step in formulating this model is to present~\eqref{E:heat_equation_2} and~\eqref{E:charge_conservation_2} in a proper dimensionless form. The current sharing temperature $T_1$ is defined by the condition $J_c(T_1)=J$.  Let us introduce a dimensionless temperature $\theta$: 
\begin{equation}
\theta=\f{T-T_1}{T_c-T_1}.
\end{equation}
Then, Eq.~\eqref{E:linear_temperature_dependence} takes the form
\begin{equation}\label{E:dimensionless_linear_temperature_dependence}
J_c=J(1-\theta).
\end{equation}
Let us introduce a fraction of the total current $u$ that flows through the stabilizer
\begin{equation}
J_1=Ju; \quad J_s=J(1-u); \quad 0\le u\le1.
\end{equation}
For $\theta \le 1$ equation~\eqref{E:piecewise_field} takes the form:
\begin{equation}
E_s=\begin{cases}
(\rho_s/d_s) J(\theta-u),& u<\theta\\
0,& u\ge \theta
\end{cases}
\end{equation}
For $\theta >1$
\begin{equation} 
E_s=\f{\rho_s}{d_s}J(1-u); 
\end{equation}
Then, Eq.~\eqref{E:charge_conservation_2}  can be written in a compact form 
\begin{equation}\label{E:charge_conservation_3}
\D{}{x} \left( \lambda^2 \D{u}{x} \right) = 
u - \Gamma \max \left[ 0, \min (\theta, 1)  - u \right],
\end{equation}
where
\begin{equation}
\Gamma =\f{\rho_sd_1}{\rho_1d_s}\gg 1
\end{equation}
and 
\begin{equation}\label{E:current_transfer_length}
\lambda=\left (\f{\bar{R}d_1}{\rho_1}\right )^{1/2}
\end{equation}
is the current transfer length which determines the length scale of the current exchange between the superconductor and stabilizer.

Taking into account\eqref{E:charge_conservation_1}, the heat sources $Q$ defined by Eq.~\eqref{E:integrated_internal_heat_sources} take form 
\begin{equation}
Q=\f{\rho_1J^2}{d_1}u^2 +\bar{R}J^2\left (\D{u}{x}\right )^2
+ J^2(1-u) \left[ \f{\rho_1}{d_1}u-\D{}{x}\left( \bar{R}\D{u}{x}\right) \right].
\end{equation}

We will measure the distances in units of thermal diffusion length $l_T$ and time in units of $\gamma^{-1}$, where
\begin{equation}\label{E:thermal_diffusion_length}
l_T =(D_T/\gamma)^{1/2};\;\;\gamma =\rho_1 J^2/d_1C\Delta T.
\end{equation}
Here $D_T = K/C$ is the effective in-plane thermal diffusivity of the conductor, $\Delta T\equiv T_c-T_1$, and the increment $\gamma $ determines the characteristic time required for the Joule heat generated in the stabilizer to warm the conductor by temperature $\Delta T$. 

In dimensionless variables, Eqs. \eqref{E:heat_equation_2} and~\eqref{E:charge_conservation_3} take the form

\begin{align}
&
\begin{aligned}
\D{\theta}{\tau}-\D{^2 \theta}{\xi^2} &= u+r\left(\D{u}{\xi}\right)^2 -(1-u)\D{}{\xi}\left ( r\D{u}{\xi}\right )\\ 
& \quad -\kappa_{\perp} (2\theta (\xi ) -\theta (\xi-2\ell) -\theta (\xi+2\ell));\\
\end{aligned}
\label{E:heat_equation_3}\\
&\D{}{\xi} \left( r \D{u}{\xi} \right) = u - \Gamma \max \left[ 0, \min (\theta, 1)  - u\right].
\label{E:charge_conservation_4}
\end{align}
with $\tau =\gamma t$, $\xi =x/l_T$ ($-\ell N \leq\xi\leq \ell N$), and $\ell=L/l_T$. 
Here 
\begin{equation}
\kappa_{\perp}=\f{K_{\perp}\Delta Td_1}{\rho_1 J^2}.
\end{equation}
Notice that Eq. ~\eqref{E:heat_equation_3} does not depend on the concrete form of the constituent relationship between electric field and current density in the superconductor. The specifics of the constituent relationship enters only in the charge conservation condition given by~\eqref{E:charge_conservation_2} and its dimensionless versions~\eqref{E:charge_conservation_3} and~\eqref{E:charge_conservation_4}.

The relative role of the interface resistance is determined by the parameter 
\begin{equation}\label{E:interface_resistance_parameter}
r=\f{\lambda^2}{l_T^2}=\f{\bar{R}}{R_0};\;\;R_0=\f{\rho_1 l_T^2}{d_1}=\f{K\Delta T}{J^2}.
\end{equation}
The results will not depend on the value of $\Gamma $ as long as $\Gamma\gg 1$. Hereafter, for the purpose of numerical solution, we take $\Gamma =10^2$. 

The main purpose of our analysis is to determine how the peak temperature inside the coil changes with the electric potential drop across the coil which is typically used to detect the normal zone. As a rule of thumb, the length of the normal zone, however defined, is considered to be a good proxy for the potential drop. Hereafter we will use a more accurate definition of the potential drop $V$ across the coil due to developing quench: 
\be\label{E:voltage_drop}
V(t)=\int_{-NL}^{NL} \f{\rho_1}{d_1}J_1(x,t)dx\approx \f{\rho_1J}{d_1}l_T\int_{-N\ell}^{N\ell}   u(\xi ,t )d\xi.
\ee
For simplicity, in this analysis we will neglect the temperature dependence of the resistivity.  The characteristic voltage
\be
V_0=\f{\rho_1J}{d_1}l_T 
\ee
determines the potential drop due to NZ with the length of the order of the diffusion length.

\section{\label{sec:model}Specific example: material properties of coated conductors at $4.2\;K$
\protect}
The solutions of Eqs. (20) and (21) provide a universal picture of quench propagation, independent of the values of the material parameters. To translate these results to a case of concrete materials and operating temperature we need to calculate the characteristic constants, such as dimensionless parameters $r$ and $\kappa_{\perp}$ as well as the increment $\gamma$ and the diffusion length $l_T$. 
One of the promising near future applications of coated conductors is magnet coils operating at or near liquid helium temperature\cite{Sch}. As an example we will use the following values of the material constants at or near $4\;K$ \cite{Ekin}:\\
Copper stabilizer: 
\bea 
\rho_1\approx 1.5\times 10^{-8}\Omega\;cm; &\\ \nonumber
k_1\approx 6.3\; W/cm K;\\ \nonumber 
c_1\approx 8.1\times 10^{-4}\;J/cm^3 K;\\ \nonumber 
d_1=40 \mu m.\nonumber
\eea
Substrate (Hastelloy):
\bea
\rho_2\approx 123\times 10^{-6}\Omega\;cm; &\\ \nonumber 
k_2\approx 8\times 10^{-4}\;W/cm\;K; (k_2=LT/\rho_2);\\ \nonumber 
c_2\approx 9\times 10^{-3}\;J/cm^3 K;\\ \nonumber 
d_2=50\;\mu m.\nonumber
\eea
The substrate thermal conductivity was estimated from the value of resistivity using Wiedemann-Franz relationship (shown in parenthesis) with $L=2.45\times 10^{-8}\;W\Omega /K^2$. 

Insulation (Polyimide $Kapton^{TM}$):
\bea
c_{ins}\approx 1.1\times 10^{-3}\; J/cm^3 K;\\ \nonumber
k_{ins}\approx 1.1\times 10^{-4}\; W/cm K;\\
d_{ins}\approx 25\mu m.\nonumber
\eea

The effective thermal in-plane conductance and the heat capacity are given by
\bea
C=c_1d_1+c_2d_2+2c_{ins}d_{ins}\approx 5.3\times 10^{-5}\;J/cm^2 \;K\; \\ \nonumber
K=k_1d_1+k_2d_2\approx 2.5\times 10^{-2}\; W/K.
\eea
It is noteworthy that at liquid helium temperature the main contribution to the in-plane heat conduction is provided by copper, but the main contribution to the conductor heat capacity is provided by the substrate. This creates a possibility of a substantial temperature difference between the stabilizer and substrate at the front of the propagating NZ. The model here is based on an assumption that the heat conduction of the buffer between YBCO and the substrate is high enough, so that such temperature difference between the substrate and stabilizer can be neglected. The in-plane thermal diffusivity
\be
D_T= K/C \approx 470\;cm^2/s.
\ee

For the sake of estimate we take the critical current density at the operating temperature $T_0$ equal to $J_c(T_0)=300\;A\;cm^{-1}$ and the transport current density $J=150\;A/cm$. Then the current sharing temperature $T_1=(T_c+T_0)/2$. Hereafter we take $T_0=4.2\;K$ and $T_c\approx 90\;K$, so that $T_1\approx  47\;K $ and $\Delta  T =T_c-T_1 =47\;K$. Then, the increment $\gamma$ (Eq. (19)) is equal to
\bea
\gamma\approx 34\;s^{-1},
\eea
while the thermal diffusion length 
\be
l_T =(D_T/\gamma)^{1/2}=\f{Kd_1\Delta T}{\rho_1J^2}\approx  3.7\;cm.
\ee
It should be noted that the estimates of $l_T$ and $\gamma$ based on the assumption that the material parameters are temperature independent with the values equal to those at $T=4\;K$  tend to produce a significant quantitative error. The resistivity and thermal conductivity of copper vary strongly within the superconducting range of temperatures and beyond it. Our purpose here is simply to illustrate the effect of the contact resistance keeping other material parameters unchanged. 
The heat conductance across the insulation $k_{\perp}$ can be estimated as
\be
K_{\perp}=\f{k_{ins}}{2d_{ins}}\sim 2.2\times 10^{-2}\;Wcm^{-2}K^{-1}.
\ee
Correspondingly, the dimensionless coupling constant (Eq. (22)) is given by
\be
\kappa_{\perp}\approx 12.
\ee
The large value of the dimensionless heat transfer constant $\kappa_{\perp}$ is noteworthy. It indicates very effective heat redistribution between the turns of the coil. It should be noted that in the case of an individual linear conductor cooled from the surface the condition of cryostability is $\kappa_{\perp}>\kappa_c < 1$\cite{PRE}. In such a conductor the condition of cryostability means that the heat generated in the stabilizer cannot maintain its temperature above the critical temperature. In an overall adiabatic coil the large value of $\kappa_{\perp}$ indicates rapid heat propagation across the insulation. 
This large thermal coupling between the turns of the coil is due to low resistance of the stabilizer, rather than a particularly high thermal conductivity of the insulation. The Joule heat generated in the stabilizer is small so that even relatively small temperature difference across the insulation $\delta T\approx \Delta T/\kappa_{\perp}$ is sufficient to transfer a large part of the generated heat flux to the neighboring sections. Also should be noted that it is the electric insulation that provides the main thermal resistance to the heat exchange between the turns, not the substrate. Indeed, $k_2/d_2\sim 16\times 10^{-2}\;Wcm^{-2}K^{-1}$, while $ K_{\perp}\sim 2.2\times 10^{-2}\;Wcm^{-2}K^{-1}$ (Eq. (33)). This is the reason why our approximation of uniform across the thickness of the conductor temperature is well justified. 
Below we will consider two different values of the heat transfer constant. One is given by Eq. (34) and the other, much smaller, $\kappa_{\perp} = 0.2$. This value corresponds to greater thickness of the insulator and/or greater resistance of the stabilizer. 

The characteristic contact resistance defined by Eq. (23) is given by
\be
R_0\sim 5.1\times 10^{-5}\Omega\;cm^2.
\ee
The characteristic voltage, Eq. (25),
\be
V_0\approx 2\;mV.
\ee

\section{\label{sec:introduction}Results\protect}
The system of equations~\eqref{E:heat_equation_3} and~\eqref{E:charge_conservation_4} were solved numerically by using an IMEX Crank-Nicolson/Adams-Bashforth method in conjunction with a fixed-point method to solve the Poisson equation with a nonlinear source term.  The purpose of the numerical solutions is to determine how the voltage drop across the coil ( Eq. ~\eqref{E:voltage_drop}) changes with time for different values of the contact resistance, how the peak temperature inside the NZ changes with time, and, by combining these results, how the peak temperature inside the coil changes with the voltage drop across the coil.  

\begin{figure*}
\includegraphics{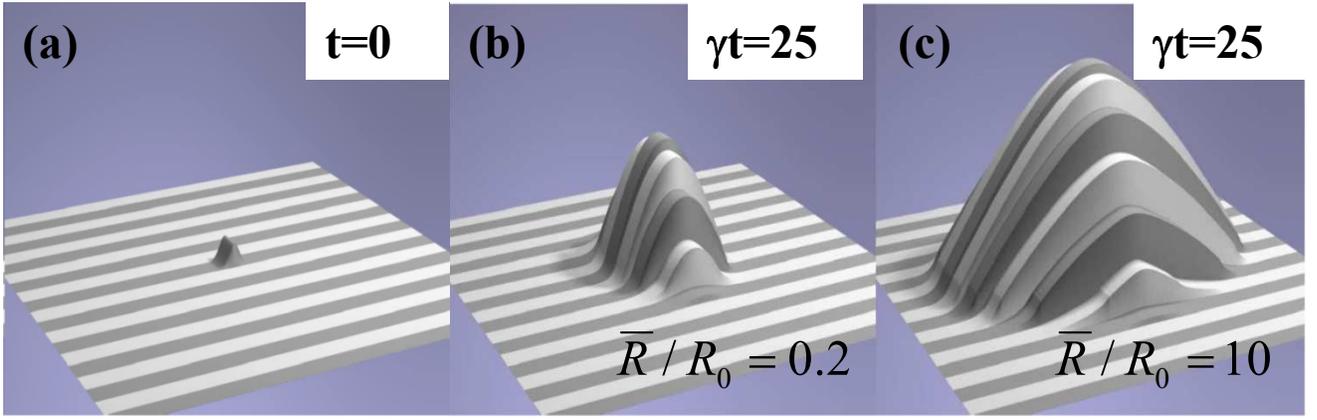}
\caption{\label{fig:wide} Temperature profiles (vertical axis) in the stacks of coated conductors obtained from the solutions of Eqs. (20) and (21). Each stripe in the figure, of either coloring, corresponds to an individual coated conductor like the ones shown in Fig. 1(b). The insulation layers between the conductors are not shown. (a) The initial condition -- a temperature perturbation in the center of the stack that gives rise to NZP. (b) The temperature profile after certain elapsed time in the stack composed of conductors with low contact resistance. (c) The temperature profile after the same elapsed time in the stack composed of conductors with higher contact resistance. The high contact resistance leads to faster expansion of the NZ. It also leads to higher peak temperature. In both cases the thermal coupling is characterized by $\kappa_{\perp}=0.2$.}
\end{figure*}

The solutions $\theta (\xi, \tau )$ presented below correspond to the initial condition in the form of a Gaussian in the center of one of the conductors in the middle of the stack.
\begin{equation}\label{E:initial_conditions}
\theta (\xi, 0)=(a-\theta_0)\operatorname{e}^{-\xi^2/2\delta^2}+\theta_0. 
\end{equation}
In all examples shown below we keep $\delta = \sqrt 2$. The peak temperature of the initial perturbation given by the constant $a$ determines whether the nucleus of the normal zone will trigger its expansion, or the extra heat associated with the initial perturbation will dissipate throughout the stack without triggering NZP. Since we choose, as an example, the transport current equal to  $50\%$ of the critical current, $J_c(T_0)=2 J$, the dimensionless operating temperature, see Eq. (11), 
\be
\theta_0=1-\f{J_c(T_0)}{J}=-1.
\ee

For~\eqref{E:charge_conservation_4} we used the following boundary conditions:
\begin{equation}
u(\ell N)=u(-\ell N)=0.
\end{equation}
The results reported below do not depend on the exact form of the boundary conditions because they are obtained from the solutions of Eqs. ~\eqref{E:heat_equation_3} and~\eqref{E:charge_conservation_4} for which the regions of elevated temperature do not yet extend to the boundaries of the stack shown in Fig. 1.

\subsection{\label{sec:NZP} Insulation with low heat conductance ($\kappa_{\perp}=0.2$).}
Figure 2 illustrates the differences between the growth of the normal zone in two stacks of coated conductors. Figure 2(a) shows the initial condition, Eq. (37), a temperature spike in the center of the stack. The stripes, both white and gray, indicate the individual conductors. The alternating coloring of the stripes allows for a better visual resolution. Insulation between the conductors is not shown. The vertical axis is the local temperature of a conductor. Figure 2(b) shows the temperature distribution in the stack composed of coated conductors with low contact resistance, $\bar{R}/R_0=0.2$, characteristic of the currently manufactured coated conductors \cite{L1}.  Figure 2(c) shows the temperature distribution after the same elapsed time in a stack composed of  coated conductors with much higher  contact resistance $\bar{R}/R_0=10$. All other parameters of the model remain the same. Notice the substantial temperature differences between the neighboring conductors. This is the result of relatively weak thermal coupling $\kappa_{\perp}=0.2$.

These results illustrate the apparent inherent contradiction laying at the foundation of the concept of high contact resistance. In the stack made out of coated conductors with high contact resistance the normal zone propagates manifestly faster, both in lateral and transverse directions. However, the amount of heat generated at the front of the propagating NZ is also higher. As a result, other things being equal, over the same period of time in adiabatic conditions the peak temperature rises higher in the stack of conductors with high contact resistance. 

However, it is important to emphasize that neither the speed of NZP alone, nor the rate of increase of the peak temperature alone can be considered as the good measures of the quench protection quality of the conductors. If the detection of the quench is based on the voltage that builds up across the coil, the higher propagation speed of NZ leads to earlier detection, so that the peak temperature by the time of detection may in fact be lower in the stack made out of coated conductors with high contact resistance. The rate at which the peak temperature increases with increasing voltage drop $dT_{peak}/dV$ is the relevant measure of the quench protection quality of the conductors. It is desirable to make this rate as small as possible, because it leaves more time for the quench protection system to be engaged without the danger of overheating the conductor.   
\begin{figure}
\includegraphics{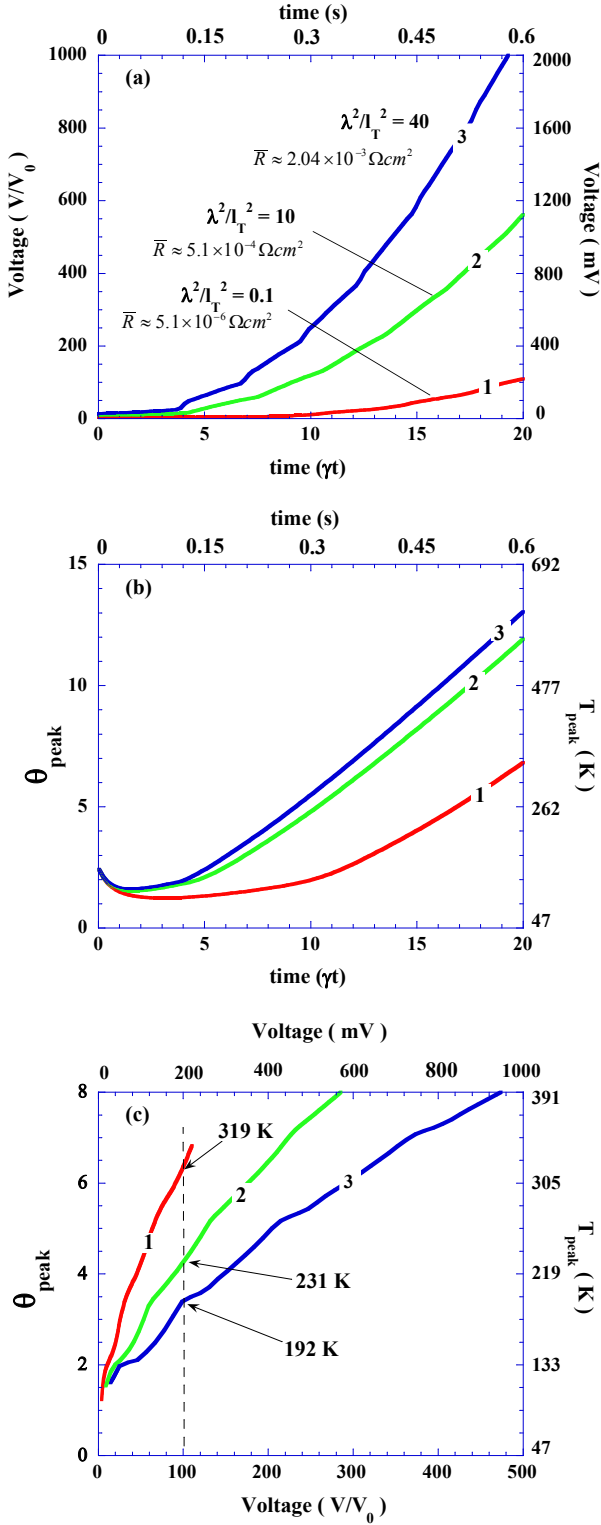}
\\
\caption{\label{fig:} The results of the simulation of NZP for three different values of the contact resistance. Curve $(1)$ corresponds to the lowest contact resistance, curve $(3)$ to the highest. (a) The voltage drop (Eq. (24)) vs time. The scale on the right and the upper scale correspond to $V_0=2\;mV$ and $\gamma =33\;cm^{-1}$ respectively. (b) The peak temperature in the center of NZ vs time. The scale on the right corresponds to the set of operating parameters shown in Eq. (40). (c) A combination of the data shown previously –- the peak temperature vs voltage drop. The dashed line illustrates that when the voltage drop reaches a certain threshold, at which point the quench protection system may be engaged, the peak temperature is the lowest in the stack made out of conductors with the highest contact resistance. }
\end{figure}

Figure 3 presents the quantitative results obtained by solving Eqs. (20) and (21). In Fig. 3(a) the curves enumerated $(1), (2)$, and $(3)$ show how the voltage drop across the coil, Eqs. (24) and (25), changes with time. The curve $(1)$ corresponds to the lowest contact resistance (see Eq. (23)) $\lambda^2/l_T^2=0.1$. The value of the contact resistance $\bar{R}=5.1\times 10^{-6}\Omega\;cm^2$ corresponds to $R_0=5.1\times 10^{-5}\Omega\;cm^2$ (Eq. (35)) and it is given here for illustration purpose only. To the best of our knowledge, the value of the contact resistance in currently manufactured coated conductors at $T=4.2\;K$  has not been measured. At $T=77\;K$ the contact resistance is much smaller than $R_0$ \cite{Polak}. The curves $(2)$ and $(3)$ correspond to progressively greater contact resistance. The scale on the right and the upper scale correspond to $V_0=2\;mV$ and $\gamma \approx 33\;s^{-1}$, Eqs. (36) and (31).  

An approximately quadratic time dependence of $V(t)$ indicates that over the time interval shown in the Fig. 3 the NZ expands simultaneously in the lateral and transverse directions.  It is evident that the initiation of the NZ growth in the outermost conductors of the stack takes place due to the transverse heat flux. The same conclusion was drawn in Ref. \cite{Sumption}. The characteristic steps in $dV(t)/dt$, most evident in curve $(3)$, correspond to the time instances when the two new outermost conductors (symmetrically located relative to the center of NZ) reach temperature $T_1$ and start generating voltage.   

Figure 3(b) shows the dimensionless peak temperature inside the stack (at the point of NZ origination, Fig. 2(a)) as a function of time. For illustration purposes the scale on the right shows the peak temperature in absolute units for the set of material and operational parameters chosen in the text as an example
\be
T_{peak}=T_1+\theta_{peak}(T_c-T_1); \;T_1= T_c-T_1=47\;K.
\ee
It is important to note that the temperature difference between the curves $(2)$ and $(3)$ is relatively small even though curve $(3)$ corresponds to four times greater contact resistance. In the conductor like the one shown in Fig. 1 the excess of heat is generated only at the front of the propagating NZ where the current exchange between the superconductor and stabilizer takes place across the interface between them. Behind the front, the current flows only through the copper stabilizer generating the minimum amount of heat determined by the value of current and the stabilizer resistance. The peak temperature of the initial condition has to be high enough to trigger NZ propagation. As the result, the initial phase of the time evolution is characterized by the falling temperature in the center of the stack due to the energy outflow from the point of NZ initiation. 

Combining the results shown in the Figs. 3(a,b) we obtained the dependence of the peak temperature in the stack on the potential drop across the coil. Here the benefits of the increased contact resistance become obvious. The slowest temperature rise with increasing voltage drop is attained in the stack composed of the conductors with the highest contact resistance. As an illustration, the vertical dashed line corresponds to a presumed threshold which triggers the quench protection system ($V/V_0=100;\; V=200\;mV$). At this level of potential drop the peak temperature inside the stack is substantially lower for the conductors with the highest contact resistance because the threshold value is reached much sooner.  

\subsection{\label{sec:NZP} Insulation with high heat conductance ($\kappa_{\perp}=12$).}

The second set of data shown in Fig. 4 corresponds to a substantially higher coupling constant $\kappa_{\perp}=12$ (Eq. (34)). The large coupling does not necessarily mean some anomalously high thermal conductivity of the insulation. It can be a result of low resistance of the stabilizer. Since the resistivity of copper strongly changes with temperature, the real thermal coupling between the turns of a coil changes somewhere within the range $0.2\leq \kappa_{\perp}\leq 12$. In Fig. 4 all notations and the values of contact resistance are the same as in Fig.3.  The only difference is the value of the thermal coupling between the neighboring conductors in the stack. The higher thermal coupling leads to a higher speed of NZ propagation in the transverse direction. The normal zone cuts very rapidly through all 21 conductors in the stack that were considered in the simulation. After that, the growth of NZ continues only in the lateral direction with a constant speed, which is why the time dependence $V(t)$ in Fig. 4(a)is mostly linear, as opposed to quadratic in Fig. 3(a).   

The temperature of the initial perturbation has to be much higher than that in the case of $\kappa_{\perp}=0.2$ in order to trigger the NZ propagation. This is why in Fig. 4(b) the initial temperature drop is more pronounced than that in Fig. 3(b). Figure 4(c) shows the same overall result as Fig. 3(c) . The peak temperature in the stack made out of conductors with the highest contact resistance is the lowest for a given value of the voltage drop. The dashed line illustrates the peak temperature for a presumed threshold $V/V_0=300$ ($V=600 \;mV$).  

\begin{figure}
\includegraphics{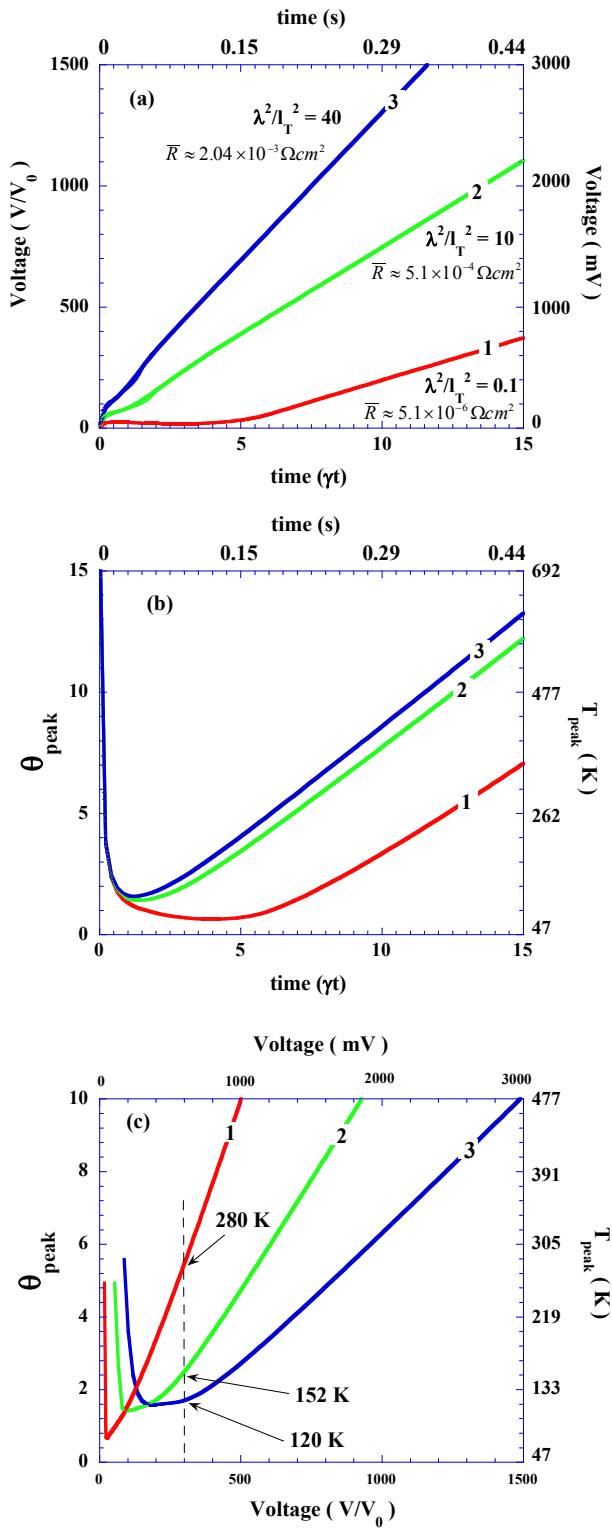}
\\
\caption{\label{fig:} The results of the simulation of NZP for three different values of the contact resistance and the thermal coupling $\kappa_{\perp}=12$. The curve $(1)$ corresponds to the lowest contact resistance, the curve $(3)$ to the highest. (a) The voltage drop (Eq. (24)) vs time. The scale on the right and the upper scale correspond to $V_0=2\;mV$ and $\gamma =33\;cm^{-1}$ respectively. (b) The peak temperature in the center of NZ vs time. The scale on the right corresponds to the set of operating parameters shown in Eq. (40). (c) A combination of the data shown previously –- the peak temperature vs voltage drop. The dashed line illustrates that when the voltage drop reaches a certain threshold, at which point the quench protection system may be engaged, the peak temperature is the lowest in the stack made out of conductors with the highest contact resistance}
\end{figure}

\section{\label{sec:summary}Summary and Rumination over Quench Protection\protect}

Here we have extended the previous treatment of the quench in coated conductors \cite{ASC,L1} to a case of a stack of conductors imitating a part of a pancake coil. Our criterion of the quench protection quality of a conductor is the rate at which the peak temperature inside the normal zone increases with the voltage drop across the coil. The smaller this rate is, the more reliable the quench protection system can be made. The detection of the quench is based on the voltage monitoring. To avoid the false alarms the threshold of triggering the quench protection system has to be made high enough. Therefore, it is highly desirable that at the moment when the threshold potential is reached the peak temperature inside NZ was as low as possible, certainly well below the temperature above which the conductor may suffer an irreversible damage. The results shown in Figs. 3 and 4 clearly indicate that the increased contact resistance leads to a faster build up of the voltage drop which more than compensates for the increased heat dissipation at the front of the propagating NZ.   
   
The origin of the increased longitudinal speed of NZP (along the conductor) was explained in \cite{L1}. The current exchange length $\lambda$, Eq. (17), replaces the thermal diffusion length as the dominant length scale that determines the NZP speed. In a coil or a stack like the one shown in Fig. 1(b) the transverse rate of NZP also increases due to increased contact resistance. The results of our simulation, as well as the experiments and   simulations presented in Ref. \cite{Sumption} show that the propagation of the NZ in the transverse direction is caused by the heat transfer across the insulation, rather than by the NZP along the conductor. The coated conductors with high contact resistance are less stable \cite{L1}, so that it takes less energy (and shorter time) to trigger NZ in the outlying sections of the coil. Therefore, in a coil the advantages of the increased contact resistance in increasing the rate of potential build up are even more pronounced than in a linear conductor cooled from the surface. This is due to a symbiotic effect of increased longitudinal NZP speed and reduced stability. 

The main apparent shortcoming of the suggested approach to improving the quench protection quality of coated conductors is that by increasing the contact resistance we artificially reduce the stability margins of the device with respect to quench\cite{L1}.  On the surface, the high stability of coated conductors is a plus. It makes quenches less frequent. However, no magnet designer will be so bold as to suggest that a coil made out of coated conductors does not need a quench protection system. The prospect of a Black Swan quench like a recent one at the Large Hadron Collider \cite{Rossi}  (unrelated to coated conductors, but noteworthy in the context of this paper) makes it inevitable that no matter how stable the current generation of coated conductors appears, a full-fledged protection system has to be designed and built. Then, what is the advantage of the high stability conductor? None - in terms of the cost of the design and construction of the quench protection system. Moreover, because of the low NZP speed in the current generation of coated conductors such a system will require a greater effort to design and validate than the state-of-the-art quench protection systems in existence today.  The low-$T_c$ superconducting magnets frequently quench when the current is ramped up for the first time. During these training sessions the weak links in the coil are revealed and healed. A very stable magnet might not quench at all for the first time it is energized with the result that a weak spot will remain undetected until it reveals itself catastrophically at some point during the lifetime of the device. 

It is obvious then, that a way to overcome these problems is to reduce the stability margins of the conductor by inserting a very thin (thermally transparent), but high resistance interface between the superconducting film and the stabilizer. Reduced stability and increased speed of NZP in such a conductor will revert the potential Black Swan quenches into already familiar, more frequent, but tamed variety.   It is possible, of course, that the existing systems of quench detection may be upgraded and refined to a level that will be sufficient for use with the current generation of coated conductors.  However, if it turns out not to be the case, then this and the previous publications\cite{ASC,L1} may be considered as a rough draft of the proverbial Plan B. 

Acknowledgment: This work was supported in part by AFOSR contract.

%\bibliography{apssamp}

.\\

\end{document}